\documentstyle[12pt,epsf]{article}
\newcommand{\be}{\begin{equation}}
\newcommand{\bea}{\begin{eqnarray}}
\newcommand{\eea}{\end{eqnarray}}
\newcommand{\beas}{\begin{eqnarray*}}
\newcommand{\eeas}{\end{eqnarray*}}
\newcommand{\ba}{\begin{array}}
\newcommand{\ea}{\end{array}}
\newcommand{\ee}{\end{equation}}

\newcommand{\tr}{{\rm Tr}\ }

\newcommand{\nbox}{{\,\lower0.9pt\vbox{\hrule \hbox{\vrule height 0.2 cm \hskip
0.2 cm \vrule height 0.2 cm}\hrule}\,}}

\newcommand{\pa}{\partial}

\newcommand\text[1]{\rm #1}

\newcommand{\CC}{\hbox{\xiiss C\kern-.4emI}}
\newcommand{\RR}{\hbox{\xiiss R\kern-.45emI}}
\newcommand{\ZZ}{\hbox{\xiiss Z\kern-.4emZ}}
\newcommand{\CCs}{\hbox{\ixss C\kern-.4emI}}
\newcommand{\ZZs}{\hbox{\ixss Z\kern-.4emZ}}
\newcommand{\pasl}{\pa\kern-.55em /}
\def\href#1#2{#2}

\begin{document}
\begin{titlepage}
\hfill
\vbox{
    \halign{#\hfil         \cr
           hep-th/0211139 \cr
            SU-ITP-02/33 \cr
           } 
      }  
\vspace*{20mm}
\begin{center}
{\Large {\bf Transverse Fivebranes in Matrix Theory}\\ }

\vspace*{15mm}
\vspace*{1mm}
{J. Maldacena,${}^1$ M. M. Sheikh-Jabbari,${}^2$ M. Van Raamsdonk${}^{2,3}$}

\vspace*{1cm}

${}^1$ {Institute for Advanced Study\\
Princeton, NJ 08540, USA}

\vspace*{0.4cm}

${}^2$ {Department of Physics, Stanford University\\
382 via Pueblo Mall, Stanford CA 94305-4060, USA}

\vspace*{0.4cm}

${}^3$ {Department of Physics and Astronomy,
University of British Columbia\\
6224 Agricultural Road,
Vancouver, B.C., V6T 1W9, Canada}

\vspace*{1cm}
\end{center}

\begin{abstract}
M-theory on the maximally supersymmetric plane wave background of eleven-
dimensional supergravity admits spherical BPS transverse M5-branes with zero
light-cone energy. We give direct evidence that the single M5-brane state
corresponds to the trivial ($X=0$) classical vacuum in the large $N$ limit of
the plane wave matrix theory. In particular, we show that the linear 
fluctuation
spectrum of the spherical fivebrane matches exactly with the set of exactly
protected excited states about the $X=0$ vacuum in the matrix model. These
states include geometrical fluctuations of the sphere, excitations of the
worldvolume two-form field, and fermion excitations. In addition, we 
propose a
description of multiple fivebrane states in terms of matrix model vacua.

Finally, we discuss how to obtain the continuum D2/M2  and NS5/M5 
theories on spheres from the matrix model. The matrix model can be viewed as
a regularization for these theories.

\end{abstract}

\end{titlepage}

\vskip 1cm
\section{Introduction}

The matrix theory conjecture \cite{bfss} states that the large N limit of the
quantum mechanics obtained from the dimensional reduction of d=10 SYM theory to
0+1 dimensions provides an exact description of light-cone M-theory in flat
eleven-dimensional spacetime. There is now a large body of evidence supporting
this conjecture (for a recent review, see \cite{wati}).

Perhaps the most basic test is that the matrix model should describe
all of the usual objects expected in M-theory. For supergravitons \cite{bfss},
membranes \cite{bfss}, and longitudinal fivebranes (fivebranes extended in the
light-cone directions) \cite{bss,clt}, the matrix model description is by now
very well known. Furthermore, it has been shown \cite{bfss,kt,tvm} that matrix
theory correctly reproduces the low-energy interactions between arbitrary
configurations of these objects expected from supergravity.

On the other hand, the matrix theory description of transverse
fivebrane states (extended in the light-cone time direction and
five transverse spatial directions) has remained somewhat
mysterious. Notably, the charge corresponding to transverse
fivebranes seems to be absent from the matrix theory supersymmetry
algebra \cite{bss}, though this does not rule out compact
transverse fivebrane states which do not carry any net charge.
There is some understanding of wrapped transverse fivebranes in
matrix theory descriptions of M-theory on tori \cite{grt, brs}.
However, we are not aware of any direct evidence for the
appearance of transverse fivebrane degrees of freedom in
non-compact matrix theory.

In this paper, we remedy this situation, providing detailed
evidence that certain quantum states in matrix theory correspond
to compact transverse fivebranes of M-theory.

Our demonstration is made possible by a number of simplifications
which result from turning on background fields corresponding to
the maximally supersymmetric plane wave of eleven-dimensional
supergravity. On the gravity side, this background permits stable
spherical transverse fivebrane states with zero light-cone energy
\cite{bmn}, which should therefore appear as vacua of matrix
theory (adding the appropriate operators to take into account
coupling to the background). On the matrix theory side, the
existence of a perturbative regime for certain values of the
background parameter $\mu p^+$ \cite{dsv1} and a powerful
supersymmetry algebra which protects energies and quantum numbers
of certain states \cite{dsv2,kp,kpark} allows us to extract exact
information about the matrix theory spectrum in the M-theory
limit. As a result, we are able to show in section 2 that the
complete linear fluctuation spectrum of a single spherical
transverse fivebrane is reproduced exactly as excited states about
the trivial vacuum in the large N limit of the matrix model. This
provides detailed support for the conjecture in \cite{bmn} that
the trivial vacuum of the matrix model corresponds to a spherical
transverse fivebrane.

In section 3, we propose and present evidence for a description of
arbitrary collections of concentric M5-branes together with
concentric M2-branes in terms of vacua in the large $N$ limit of
matrix theory. We note that at finite $N$, the distinction between
fivebranes and membranes is ambiguous. In section 4, we note a particular 
limit of the matrix model that can be used to describe the decoupled 
D2/M2 brane theories on a sphere. In section 5
we discuss similar  limits which give the IIA NS5 brane little string
theory on a fivesphere or the M5 brane theory on the fivesphere.

\section{The fivebrane spectrum from matrix theory}

Our starting point is the observation \cite{bmn} that in the
presence of a particular set of background fields, namely the
maximally supersymmetric plane wave solution of eleven-dimensional
supergravity, the classical action for the M5-brane has a zero
light-cone energy solution corresponding to a stable spherical
transverse M5-brane with radius
\be
\label{radius} r^4 = {\mu p^+ \over 6} \; ,
\ee
as shown in detail in Appendix B. We therefore
expect that matrix theory with these background fields turned on
should have a zero-energy vacuum state corresponding to the
spherical fivebrane.

The relevant matrix model was described in \cite{bmn}. The Hamiltonian is
\bea
\label{PPmatrix}
H &=& R\ \tr \left( {1 \over 2} \Pi_A^2 - {1 \over 4} [X_A, X_B]^2
- {1 \over 2} \Psi^\top \gamma^A [X_A, \Psi] \right)\cr
&& + {R \over 2} \tr
\Big(\sum_{i=1}^3 {\left({\mu\over
3R}\right)^2 } X_i^2 + \sum_{a=4}^9 \left({\mu \over 6R}\right)^2 X_a^2
\cr
&& \qquad \qquad + i {\mu \over 4R} \Psi^\top \gamma^{123} \Psi
+ i {2\mu \over 3R} \epsilon^{ijk} X_i X_j X_k \Big)\ .
\eea
This should describe M-theory on the maximally supersymmetric plane-wave background in the large $N$ limit with fixed $p^+ = N/R$.

In \cite{bmn} it was shown that this matrix model has a discrete set of
classical supersymmetric vacua given by $X^i = {\mu \over 3R}J^i$ where $J^i$
are the generators in an arbitrary N-dimensional reducible representation of
$SU(2)$. It is well known \cite{dielectric}
that such matrix model configurations correspond to
collections of membrane fuzzy-spheres with classical radii related to the
dimensions $N_i$ of the individual irreducible representations making up $J^i$
by
\[
r_i^2 = {\mu^2 \over 9 R^2} {N_i^2-1 \over 4} \; .
\]
These vacua are expected, since M-theory in the plane wave background also 
admits
stable spherical membranes with radii $r = \mu p^+ / 6$ with zero
light cone energy $- p_+=0$.

In \cite{dsv2}, it was shown that all of these vacuum states must
be exact quantum-mechanical vacua. Thus, the spherical fivebrane
state should correspond to a quantum state given by some linear
combination of these vacuum states. Since the classical fivebrane
solution sits at the origin of the three dimensional space in
which all of the membrane sphere solutions extend, a natural
candidate for the fivebrane state is the trivial $X=0$ vacuum for
which $J^i$ corresponds to $N$ copies of the trivial
representation of $SU(2)$, as conjectured in \cite{bmn}. The
situation is similar to the one in \cite{jpms}; in Appendix D we
make this connection a bit more precise.

\subsection{Perturbation theory}

The expansion parameter in perturbation theory for the matrix
model about the $X=0$ vacuum is \cite{dsv1}
\be 
\label{coupling}
    { N R^3 \over \mu^3} =  {  g^2_{0} N  \over \mu^3 } = {N^4 \over
( \mu p^+)^3 }\ , 
\ee 
where $g_0$ is the zero brane coupling.
At first sight, the $X=0$ state looks very
little like a spherical fivebrane. Classically, it resides at the
origin for any value of $\mu p^+ $, while the fivebrane is supposed to
have radius $r \propto (\mu p^+)^{1 \over 4}$. 
For weak coupling, it is straightforward to calculate that 
\be 
\label{radiusper} \bar{r}^2
\equiv \langle 0| {1 \over N} \tr(X_a^2) |0 \rangle = {18 N^2  \over
\mu p^+}(1 + {\cal O} ( N^4/(\mu p^+)^3)) 
\ee 
Thus, for fixed $N$ in the perturbative regime the size of the
$X=0$ state actually decreases as $N^2/(\mu p^+)$ when $\mu p^+ $
becomes large rather than increasing as $(\mu p^+)^{1/4}$. On the
other hand, for the validity of the matrix theory conjecture, the
classical expression (\ref{radius}) for the fivebrane radius need
only be reproduced in the limit of large $N$ with fixed $\mu$ and
$p^+ = N/R$.
 In this limit, the effective coupling (\ref{coupling})
always becomes large, so
perturbation theory is inapplicable.
Thus, it is possible  that there is a
transition from (\ref{radiusper})  to (\ref{radius}) when we take
the large $N$ limit.
Intriguingly, both  (\ref{radius}) and (\ref{radiusper}) become of the
same order of magnitude when the coupling (\ref{coupling}) is of order
one.

\subsection{Protected quantities}

From the discussion in the previous section, it appears that any fivebrane-like
properties of the $X=0$ vacuum will emerge only as strong coupling effects in
the matrix model, and would therefore be extremely difficult to observe
directly. Fortunately, as shown in \cite{dsv2}, the $SU(4|2)$ symmetry algebra
of the matrix model implies that certain physical quantities are exactly
protected for all values of $\mu > 0$. For any value of $N$, these may be
calculated in the $\mu \to \infty$ limit where the theory is free and then
extrapolated to any desired value of $\mu >0$. In this way, it is possible to
obtain reliable information about the matrix model at a given value of $\mu p^+$
even in the large $N$ limit where the theory is strongly coupled.

The protected quantities that we will be interested in are the energies and
quantum numbers of certain excited states about the $X=0$ vacuum. As discussed
in \cite{dsv2,kpark}, physical states of the matrix model must lie in representations
of $SU(4|2)$ which are comprised of finite collections of representations of the
bosonic subalgebra $SO(6) \times SO(3) \times {\rm Energy}$. Among the
physically allowable $SU(4|2)$ representations, there are certain BPS
representations which are exactly protected, that is, the energy 
(in units of
$\mu$) and $SO(6) \times SO(3)$ state content cannot change as $\mu$ is
varied.\footnote{Note that only certain BPS multiplets are exactly protected,
since others may combine and form non-BPS multiplets.} Thus, any such
representation present at $\mu = \infty$ (where the exact spectrum was
calculated in \cite{dsv1}) must be in the spectrum for any value 
of $\mu > 0$.

It turns out that the spectrum of excitations about the $X=0$ vacuum contains
infinite towers of these exactly protected representations \cite{dsv2}. To
describe these, we recall that at $\mu = \infty$, the theory becomes quadratic,
with Hamiltonian
\[
H_2 = \mu \tr({1 \over 3} A_i^\dagger A_i + {1 \over 6} A_a^\dagger A_a + {1
\over
4} \psi^{\dagger I \alpha} \psi_{I \alpha} ) \; ,
\]
and the spectrum of excitations about the $X=0$ vacuum are generated by traces
of products of the matrix creation operators
\[
\psi^\dagger_{I \alpha} ~, \qquad A^\dagger_i =
\sqrt{\mu \over 6R}X^i - i \sqrt{3R
\over 2 \mu} \Pi^i ~ ,
\qquad A^\dagger_a = \sqrt{\mu \over 12 R}X^a -  i \sqrt {3R \over \mu} \Pi^a
\; .
\]

Among the states containing only a single trace, we have a single tower of
exactly protected $SU(4|2)$ multiplets, described by primary states
\be \label{protected}
C^{a_1 \cdots a_n} \tr(A^\dagger_{a_1} \cdots A^\dagger_{a_n}) | 0 \rangle
\ee
plus states obtained from these by acting with supersymmetry generators. Here
$C^{a_1 \cdots a_n}$ is a completely symmetric, traceless tensor of $SO(6)$. The
remaining exactly protected primary states are identical in form but have a $U(N)$ index
structure involving more than one trace. In Appendix A we give a
simple argument for why these states are protected.

\begin{figure}
\centerline{\epsfysize=2truein \epsfbox{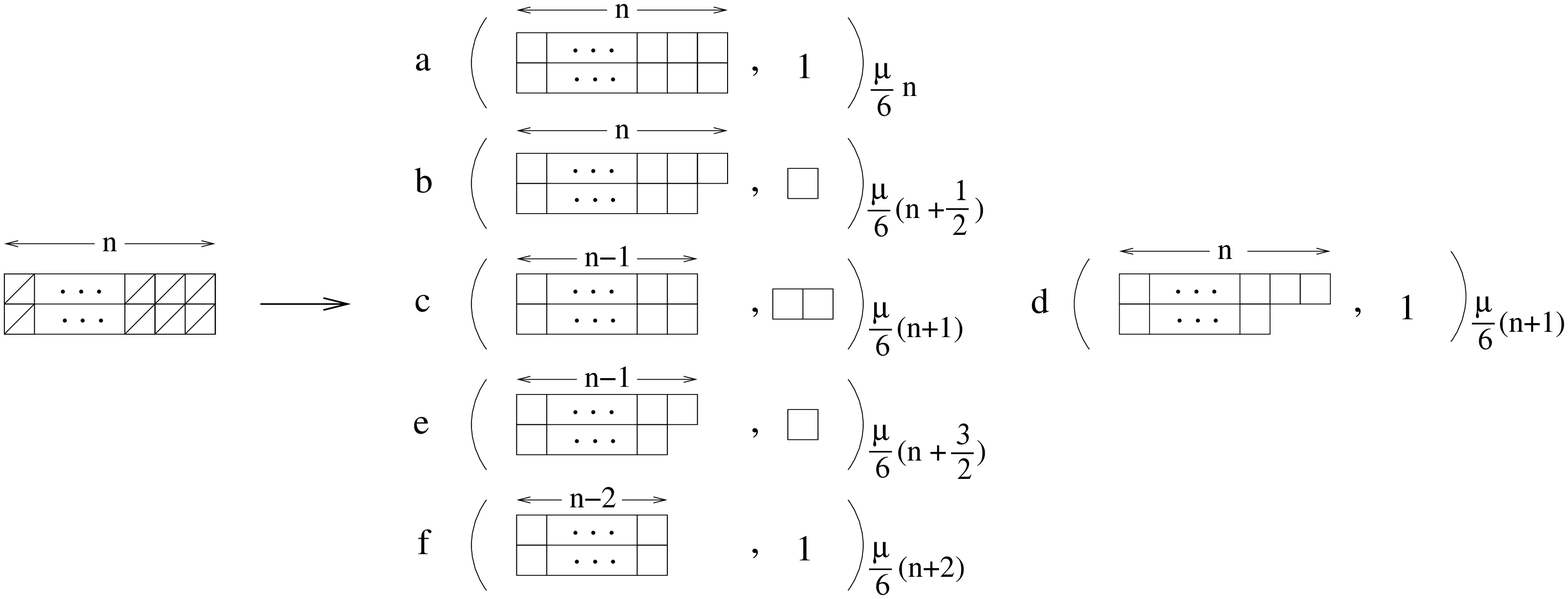}}
 \caption{$SU(4) \times SU(2)$ tableau and energies for exactly protected
excitations about the $X=0$ vacuum.}
\end{figure}

The complete set of $SO(6) \times SO(3)$ representations that
descend from these primary states is displayed in figure 1. This
shows the spectrum of single-trace exactly protected multiplets about
the $X=0$ vacuum in the large $N$ limit defining
M-theory.\footnote{For finite $N$, there are identities which
relate certain single trace operators with multiple trace
operators so we have a truncated version of the spectrum.} If the $X=0$ vacuum represents a spherical M5-brane,
these states should be among the excitations of the fivebrane.

In fact, the spectrum of exactly protected matrix theory states given in 
figure
1 matches precisely with the linear fluctuation spectrum of the spherical
fivebrane solution!

As shown in  Appendix B, geometrical fluctuations of
 the fivebrane in the radial and $x^-$ directions
are described by the representations {\bf a}
and {\bf f} in figure 1, geometrical fluctuations in the three transverse
directions give modes which make up the representation {\bf c}, two-form
fluctuations yield the remaining tower of bosonic states {\bf d}, and the
representations {\bf b} and {\bf e} are excitations of the worldvolume fermions.

In a similar way, the exactly protected multi-trace states will match with protected fivebrane fluctuations containing several quanta. 

Thus, the exactly protected excited states above the $X=0$ vacuum precisely
correspond to the fluctuations of the spherical transverse M5-brane in the
plane wave background. This represents compelling evidence that the $X=0$
vacuum of the matrix model does indeed describe the spherical
transverse fivebrane.

\section{Multiple fivebranes}

In addition to the single fivebrane state we have discussed, M-theory
on the maximally supersymmetric plane wave should contain states with
concentric fivebranes of arbitrary radii, as is the case for the
spherical membranes. We will now propose a matrix model
description of these and then present evidence for the proposal.

At finite $N$, there is a natural one-to-one correspondence between vacua of the
matrix model and ways of distributing $N$ units of momenta between any
number of membranes. A vacuum corresponding to a partition $N = N_1 + \cdots N_m$ (where $N_i$ label the sizes of the associated $SU(2)$ irreps) has a membrane interpretation as concentric fuzzy spheres
with radii proportional to the individual momenta $N_i/R$. On the
other hand, we expect equally many fivebrane states, since states with $N$ units of momentum divided between $k$ fivebranes would also be labelled by partitions of $N$.\footnote{As discussed below, we might also have vacua which include both membranes and fivebranes.}

Since we have already associated all the vacua with membrane
states, it is evident that the distinction between membrane states and
fivebrane states must be somewhat ambiguous at finite $N$. As an
example, we note that the state corresponding to a partition $N=1 +
\cdots + 1$ would be given a membrane interpretation
as $N$ membranes each carrying one unit of momentum, but this is
precisely the state that we have associated with a single fivebrane
(carrying $N$ units of momentum).

In fact, there is a natural dual fivebrane interpretation for each of the
vacuum states. To describe this, note that any
partition of $N$ may be represented by a Young diagram whose column
lengths are the elements in the partition. In the membrane
interpretation, such a diagram would correspond to a state with one membrane for each column with the number of boxes in the 
column corresponding to the number of units of momentum. In the dual
fivebrane interpretation, it is the rows of the Young diagram that
correspond to the individual fivebranes, with the row lengths
corresponding to the number of units of momentum carried by each
fivebrane, as shown in figure 2. 

With this interpretation, it is clear that the number of fivebranes is equal to the size of the largest irreducible representation, while the momentum $M_n$ carried by the $n$th fivebrane is equal to the number of irreducible representations with size greater than or equal to $n$.

\begin{figure}
\centerline{\epsfysize=1.5truein \epsfbox{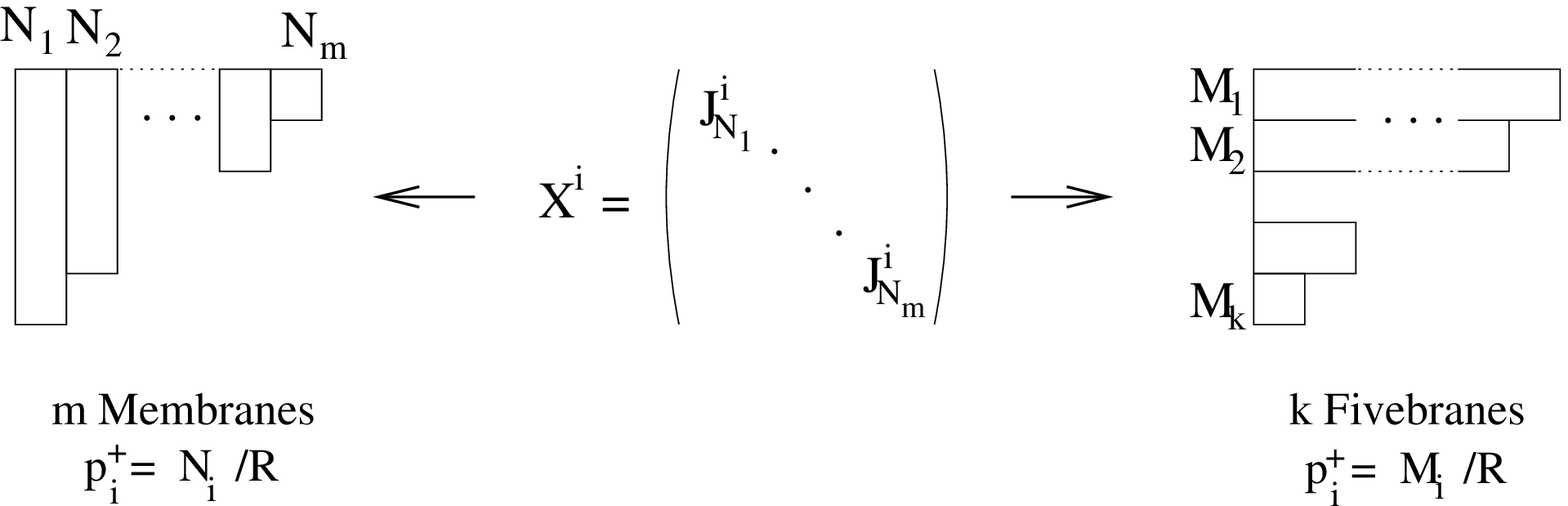}}
 \caption{Dual membrane and fivebrane interpretations for general vacua.}
\end{figure}

Since we now have both membrane and fivebrane interpretations for
each of the vacua, an obvious question is which of these
interpretations is more appropriate.\footnote{In principle, we may
answer this question by finding the expectation value of various
operators corresponding to multipole moments of the charge
distributions to determine the geometry of the state.} In general,
the answer will depend on the values of $(\mu p^+)$, $N$
and  the parameters describing the vacuum of interest.
Given any fixed choice of vacuum, the membrane interpretation will
be correct for small values of the effective coupling (i.e. sufficiently small values of $1/(\mu p^+)$), since the
theory becomes free in this limit and the classical geometry
corresponding to concentric fuzzy spheres will not receive quantum
corrections. On the other hand, our single fivebrane example
suggests that if the number of representations is large enough
so that the coupling is large,
 then the fivebrane interpretation should be
appropriate. For general intermediate values of the parameters,
the identification as membranes or fivebranes is likely ambiguous.

The situation is  clearer in the large $N$ limit defining
M-theory. Here, to define a state with a fixed number of membranes
with various momenta, we keep the number of irreducible
representations and the ratio of their sizes fixed as we take the
large $N$ limit. The sizes $N_i$ of the individual representations
thus go to infinity and the momentum fractions $N_i/N$ carried by the
various membranes become continuous parameters. On the other hand, to
define states with fixed numbers of fivebranes, we keep the sizes of
representations fixed and take the number of representations to
infinity. In this case, the number of fivebranes corresponds to the
maximum representation size, and the momentum fraction carried by the
$n$th fivebrane is $M_n/N$ where $M_n$ is the length of the $n$th row in the Young diagram. 

In terms of the Young diagrams of figure 2, membrane/fivebrane states
correspond to diagrams with a fixed number of columns/rows in the
large $N$ limit. One may also consider more general limits in which both the
number of rows and the number of columns become infinite. It is
natural to identify states with $m$ infinite columns and $k$ infinite
rows with configurations including both $m$ concentric membranes and $k$
concentric fivebranes. On the other hand, it is not clear how to
interpret states where the number of infinite rows or columns becomes
infinite (e.g. the large $N$ limit of the diagram with rows of length
$1,2,3, \dots N$ or the square diagram with N rows of length N).

Before proceeding to give evidence for this proposal, we note that the
description of multiple fivebrane states is very different from the
usual classical picture in which different blocks in block-diagonal
matrices correspond to different objects. For example, it is not true
that taking a block diagonal configuration where each block represents
the classical matrix for a single fivebrane leads to a multiple
fivebrane state.

\subsection{Evidence for the proposal}

Given this explicit proposal for the identification of multiple
M5-branes states with matrix model vacua in the large $N$ limit, we
would now like to see what evidence supports it.

Firstly, with our construction, there is a one-to-one correspondence between
the ways of dividing $N$ units of momenta between $k$ fivebranes and
matrix model vacua involving irreducible representations of size $k$.

Further evidence comes by considering the exactly protected states about the
various vacua, as we have done for the single fivebrane case. Consider
first the M-theory states with $k$ coincident spherical
fivebranes. The corresponding vacua are those with $N$ copies of the
$k$ dimensional irreducible representation in the large $N$ limit.
From the results of \cite{dsv1,dsv2}, it is straightforward to verify
that the exactly protected excited states above these vacua are the same for
any value of $k$ (at any fixed $N$ or in the large $N$
limit).\footnote{Here, we mean states which are exactly protected by
$SU(4|2)$ representation theory. As we discuss below, there may be
other states whose energies are protected for other reasons even
though group theory does not forbid them from receiving energy
shifts.} This is consistent with the idea that the $k$ coincident
fivebranes form a bound state whose protected fluctuations are the
same as for a single fivebrane.

Next consider the general vacua corresponding to partitions
\[
(1,\dots,1,2,\dots,2,\dots,k,\dots,k)
\]
in the limit where the number
of each representation goes to infinity with fixed ratios. Here, the
single-trace exactly protected states form $k$ copies of the
single-trace exactly protected spectrum of a single fivebrane. This
fits in well with our interpretation of such a state as $k$ concentric
fivebranes at different radii, since the $k$ copies may be interpreted as
independent fluctuations of the $k$ individual fivebranes. In the case
where the fraction of representations of size $n$ drops to zero for some $n<k$, we lose one set of protected states, and this may be interpreted as the
$n$th largest fivebrane shrinking to form a bound state with the
$(n+1)$st largest fivebrane.

It is important to note that the calculated fluctuation spectrum
in the Appendix did not depend on the radius of the fivebrane.
This is the reason that we simply have $k$ copies of the
single-fivebrane fluctuation spectrum for the matrix model states
corresponding to $k$ fivebranes at arbitrary radii. For the
$k$-coincident fivebrane state, it is possible to obtain some
evidence that the corresponding matrix model state has the correct
physical size by repeating the suggestive calculation of section
2. For the state with $m$ copies of the $k$ dimensional
representation ($N=mk$) we find
\[
\bar{r}^2 \equiv \langle {1 \over N} \tr(X_a^2) \rangle = 18 \left( {R \over \mu}
\right) m (1 + {\cal O}(m R^3/\mu^3))  \; .
\]
As $m$ is increased with fixed $\mu p^+ = \mu mk/R$, both the size of the
state and the size of the perturbative corrections grow, and the size
matches with the expected value $\bar{r} \approx (\mu p^+ / k)^{1/4}$ just as
the effective coupling becomes of order 1. Of course, there is no
reason to expect that such a calculation should give the desired
radius (which we only require to be reproduced for $m \to \infty$)
except that it worked for the single fivebrane case.

\subsection{Chiral operators for multiple membranes and fivebranes}

To give a final piece of evidence that the proposed description of
fivebranes is correct, we consider in some detail the matrix model
vacua with $m$ copies of the $k$ dimensional irreducible
representation. Depending on the parameters, this vacuum may describe
$m$ coincident membranes (e.g. if $k \to \infty$) or $k$ coincident
fivebranes (e.g. for $m \to \infty$). In general, the excitations
about this vacuum state (at large $\mu$) are generated by $m \times m$ matrix
oscillators in the $SU(4|2)$ representations corresponding to
single-column supertableau with $2,4, \dots, 2k$ boxes.

The exactly protected single trace states arise from a trace
containing up to $m$ two-box oscillators (traces of more than this
number are not independent). For $m \to \infty$, we obtain the
complete spectrum of figure 1, and as we have discussed, these states
correspond to protected fluctuations of the bound state of $k$
fivebranes. On the other hand, in the membrane limit $k \to
\infty$ with fixed $m$, we still have a truncated version of the
spectrum in figure 1, and we may ask how this should be interpreted in
the membrane picture.

To answer this, recall that the $SO(8)$ superconformal theory
describing the low-energy theory of $m$ coincident M2-branes in
M-theory has chiral operators in symmetric traceless representations
of $SO(8)$ with up to $m$ indices (analogous to the chiral operators
of $N=4$ SYM with in symmetric traceless representations of $SO(6)$.)
These operators should correspond to states of the membrane theory on $S^2
\times R$. In the plane wave background, the theory living on the
spherical membranes  only has an $SO(6)$ subset of the $SO(8)$
R-symmetry preserved. Therefore, if states corresponding to the chiral operators
survive for spherical M2-branes in this background, we would expect
them to show up as states in symmetric traceless representations of
the $SO(6)$ with up to $m$ indices. The exactly protected states we
have discussed have precisely these quantum numbers and this cutoff
(and also the right relationship between energy and quantum numbers),
so it is natural to conclude that they correspond to the chiral operators.

Given that we see evidence of the chiral operators of the $SO(8)$
superconformal theory, we can now ask whether the $k$-fivebrane
states also have excitations corresponding to the chiral operators of
the $(0,2)$ superconformal theory. In this case, the chiral primary
operators lie in symmetric traceless representations of the $SO(5)$
R-symmetry with up to $k$ indices. For spherical fivebranes on the
plane wave background, only an $SO(3)$ subgroup of this $SO(5)$ is
preserved, so in this case, we expect protected states in $SU(2)$
representations with spins up to $k$. In fact, there is a set of
single trace states in the spectrum with precisely these quantum
numbers, obtained by taking traces of the single oscillators mentioned
above.\footnote{The oscillator in the $SU(4|2)$ representation
with $2n$ vertical boxes includes ``primary'' states with purely
$SU(2)$ quantum numbers for spin $n$ plus states obtained from this by
acting with supercharges.}

We would thus like to associate these single oscillator states with chiral
operators of the $(0,2)$ theory, however it is not clear a priori that
they should be protected. These multiplets are BPS, however as
discussed in \cite{dsv1}, BPS $SU(4|2)$
multiplets corresponding to supertableau with more than three rows (or
more than one box in the third row) have the possibility of combining
with certain other BPS multiplets to receive energy shifts. In fact,
there are examples \cite{dsv1,kp} of states in
precisely these single-column representations which receive energy
shifts in perturbation theory.

On the other hand, there is some
evidence that the particular states in question do not receive energy
shifts. For the case $m=1$, it was argued in \cite{dsv1} that the
4-box and 6-box single oscillator states cannot receive corrections to
any order in perturbation theory, while the 8-box and 10-box states
are protected at least to leading order in perturbation theory. A more general
argument is that the quadratic fluctuation spectrum of a spherical
membrane in this background gives precisely the set of $SU(4|2)$
representations corresponding to supertableaux with $2,4,6 \dots$
vertical boxes. By analogy with the fivebrane case, we would then
expect protected states in the matrix model which could be identified
with these states in the large $N$ limit. The natural candidates for
such protected states are the single-oscillator states (these are the only
choice for $m=1$). In other words, we expect that the states that
match with the membrane fluctuation spectrum in the large $k$ limit
should be protected, and that these states should correspond to the
chiral 5-brane operators in the large $m$ limit with fixed $k$.

\section{The matrix model as a regularization of the \\
D2/M2 brane theories}

Since the D0 brane matrix model expanded around vacua with large
$SU(2)$ representations looks like the D2 theory on a fuzzy sphere it
is natural to ask whether we can think of this matrix model as
a regularization of the D2 theory. It is interesting as a regularization
because it preserves the 16 supersymmetries of the D2 brane theory.
These supercharges anticommute to the Hamiltonian plus
rotations of the sphere. In the limit that the size of the sphere
is large these rotations become
 translations and rotations of the spatial plane.
On the other hand, a lattice regularization will break more
supersymmetries (see \cite{kaplan} for a discussion).
In the next subsection we discuss this point in more detail.

\subsection{Decoupling limits}

Consider the matrix model expanded around the $k$ membrane
vacuum where we have $k$ copies of the $N$ dimensional representation
of $SU(2)$.
We find that the theory looks like a fuzzy sphere of radius $\mu^{-1}$
with non-commutativity parameter and coupling constant
\be \label{dtwolimit}
 \theta  = { 1 \over \mu^2 N} ~,~~~~~~~~~
g^2_{2 YM} = { g^2_{0 YM} \over \mu^2 N}
\ee

We are interested in the limit $N \to \infty$ keeping the two
dimensional gauge coupling $g^2_{2 YM}$  and $\mu$ fixed.
In this limit, the noncommutativity parameter vanishes and we obtain a continuum theory on a two sphere with
sixteen supercharges and action
\bea 
S &=& { 1 \over g^2_{YM} }
\int dt { d \Omega \over \mu^2}\ {\rm tr} \left( -{1 \over 4} F^{\mu
\nu} F_{\mu \nu} - {1 \over 2} (D_\mu X^a)^2 - {1 \over 2} (D_\mu
\phi)^2 +{i \over 2} \Psi^\dagger D_0 \Psi 
\right. \nonumber\\
&&- {i \over 2}
\epsilon^{ijk}
\Psi^\dagger \gamma^i x^j D_k \Psi 
 + {1 \over 2} \Psi^\dagger \gamma^i x^i [\phi, \Psi] + {1 \over
2} \Psi^\dagger \gamma^a [ X^a, \Psi]
 + {1 \over 4} [X_a, X_b]^2 + {1 \over 2}  [\phi, X^a]^2 \nonumber\\
&& \left. -{\mu^2 \over 8}   X^a X^a - {\mu^2 \over 2} \phi^2 -
{3i\mu \over 8}
\Psi^\dagger \gamma^{123} \Psi +
{\mu \over 2} \phi \epsilon^{ijk}
x^i F_{jk} \right)\ ,
\label{dtwotheory}
\eea
where trace and the commutators are those of $k\times k$ matrices 
and $x_i^2=1$. The first and second line contain the terms present in
the usual $2+1$ dimensional SYM theory. The third line contains mass
terms for the scalars and fermions plus an extra $\phi F$ interaction.
The radius of the sphere is proportional to $\mu^{-1}$.
The derivation of this Lagrangian, as well as the supersymmetry
transformations, can be found in Appendix C.

In the $\mu \to 0$ limit this action becomes the action of $2+1$
Yang Mills in flat space and the supersymmetry of the theory becomes
the supersymmetry of the flat space $2+1$ Yang Mills theory.

Similarly we could consider the $g^2_{YM} \to \infty$ limit keeping
$\mu$ fixed which would give us the superconformal theory associated
to M2 branes on $S^2 \times R$.

The $\phi$ scalar in (\ref{dtwotheory})  is basically associated to the
radial direction.
We can imagine adding a magnetic flux over $S^2$.
If the theory we start with
is the
$U(N k) $ matrix model then adding a  magnetic flux  on $S^2$ is
equivalent to starting with the $U(N k + n)$ matrix model.
In this case, the final theory will contain an additional
flux $ \int_{S^2}  Tr[F] = n$. As a result, the vacuum of the theory
(\ref{dtwotheory}) is given by a $\phi \sim  n \mu/k $.
Note that if we start with an  $SU(Nk)$ matrix model we get
a two brane theory of the form (\ref{dtwotheory}) with a gauge
group $U(k)$ but with the zero modes on the sphere of the center of
mass $U(1)$ removed.

In the theory (\ref{dtwotheory}), we can also consider vacua
where $\int_{S^2}  Tr{F} =0$ but where $F$ is diagonal  with
$\phi$ similarly diagonal and different entries along the diagonal.
 For example if $k=2$ we can choose
 $F \sim \mu\phi \sim  diag(n,-n) $ . This vacuum
 has zero energy  and
can be thought of as coming from a representation of $SU(2)$ containing
two irreducible representations, one
of dimension $N + n$ and one of dimension $N- n$. Clearly this
configuration should be included in the path integral of the theory
(\ref{dtwotheory}). In fact it is possible to estimate the
tunnelling amplitude in the matrix model between this vacuum and
the vacuum with two representations of equal size.
In supersymmetric quantum mechanics the tunnelling amplitude between
two supersymmetric
vacua can be estimated as the difference in superpotential between them.
In our case the superpotential in question has the form
$ W  \sim { 1 \over g_0^2 }
Tr[ \epsilon_{ijk} X^i X^j X^k + \mu X^i X^i ]$ so that the
difference in superpotential is proportional to the difference in
the trace of the
second
casimir of the $SU(2)$ representation that defines the vacuum.
In particular, the superpotential difference between a vacuum with
two representations of size $N$ and the one with representations
of sizes $N+1$, $N-1$ is of the order of $N/g_0^2$. In our limit
we are keeping this constant so that the tunnelling amplitude is not
suppressed.
In other words, we cannot isolate a particular vacuum  of the matrix
model, but the vacua we can tunnel into have a perfectly good
interpretation from the D2 brane point of view and should be included
in the definition of the theory.
Vacua where the difference in dimension of the representations goes
to infinity as $N $ goes to infinity are very far away and the
matrix model cannot tunnel to them in finite time.

In order to understand how the action (\ref{dtwotheory}) relates to an M2-brane theory, it is useful to take a $U(1)$ gauge group in
(\ref{dtwotheory}).
We can then dualize the $U(1)$ field strength. Due to $\phi F$ coupling
in (\ref{dtwotheory}) the dualization is slightly different than
the one for the flat space D2 action. Namely the dual scalar is
defined by
\be
 d \varphi ={ 1 \over g^2} (  * F  +  \mu \phi dt)
\ee
Then the equations of motion for $\phi$ and $\varphi$ can
be rewritten as
\be \label{simpleeq}
 \nabla^2 \varphi + { \mu \over g^2} \partial_0 \phi =0 ~,~~~~~~
\nabla^2 \phi - \mu g^2 \partial_0 \varphi=0
\ee
These can be viewed as the equations of motion for two of the
transverse scalars of the $M2$ theory. More precisely, let us
denote by $Z$ a complex combination of two of those scalars.
Then take a configuration with $Z_0 = g^2 e^{ i \mu t}$ which
is classically rotating. Then we can expand to first
order $Z = (g^2 + \phi) e^{ i \mu t + i \varphi} $.
The equation of motion for  $Z$ is  of course a harmonic oscillator
equation with frequency $\mu$. This leads to the above equations
(\ref{simpleeq}) for $\phi, \varphi$. In the large $g^2$ limit, we 
see that 
 we get the M2 theory expanded around a state
with very high angular momentum in an $SO(2)$ subgroup of $SO(8)$.
For this reason we only see an explicitly $SO(6)$ symmetry in
(\ref{dtwotheory}). For the nonabelian case, one would be tempted to say that the theory we get is simply the usual M2 theory since the angular
momentum can be carried only by the overall $U(1)$.

\subsection{Gravity perspective}

Now we analyze these decoupling limits from the gravity perspective. 
The main point of this exercise is to learn how to do it since for
the NS5/M5 case we will only have this gravity description. 

We wish to take the limit in which the radius of the 
M2 brane becomes very large in Planck units. This is achieved by 
taking large values of $ r = \mu p^+ $. It is convenient to 
rewrite the relevant terms in the metric around this large radius 
two sphere that the brane is wrapping as 
\be \label{metr}
ds^2=  - 2 dx^+ dx^- -   \mu^2 r^2 (dx^+)^2  + r^2 d\Omega_2^2  + \cdots
= { 1 \over r^2 \mu^2} (dx^- )^2  -  \mu^2 r^2 (d{\tilde x}^+)^2
+ r^2 d\Omega_2^2  + \cdots
\ee
where we defined the variable ${\tilde x}^{+} = x^+ + x^-/(r \mu)^2 $
and the dots indicate terms that are not needed for our discussion. 
We will be interested in keeping $\mu$ fixed and taking $r \to \infty$.

The physical compactification radius of the $x^-$ direction is
$\tilde R = R/(\mu r)$.\footnote{Note that when we shift $x^-$ by its period
we also shift $x^+$ but the shift in $x^+$ will go to zero in the limit
that we are taking when we keep the two dimensional gauge coupling 
of the D2 theory fixed.} The D2-brane theory is then characterized by a dimensionful coupling given by the usual relation $g_{YM} = \tilde{R} /l_p^{3 / 2}$. In the worldvolume theory, it is convenient to rescale the metric so that the sphere has unit radius, after which the theory may be characterized by the dimensionless coupling 
\be
g_{YM} r^{1 / 2} = \tilde R \sqrt{r}  = {R \over r^{1/2} \mu} 
\ee
which we keep fixed in the limit (we have set $l_p = 1$).


We see that in order for this to be finite as $r \to \infty$ with fixed $\mu$, we need that $R^2/r \sim R^3/N \sim g_{0}^2/N$ is finite in agreement
with (\ref{dtwolimit}), where we have used $r \sim \mu p^+ = \mu N/R$. 

\section{The matrix model as a regularization of the \\
NS5/M5 theories}

In this case we can only use the supergravity reasoning to 
guide us on what the limit should be since we cannot derive the 
NS5 (or a non-commutative version \cite{omth} for finite $N$) 
at weak coupling from the matrix model. 
Now the relation between the radius and the momentum is 
(\ref{radius}). We can again write the metric in a 
form similar to (\ref{metr}). If we scale $R$ in a suitable way 
and $N\to \infty$ we expect to get an NS5 brane theory on a 
fivesphere of radius $r$. 

The NS5 brane theory, the so called little string theory, is 
characterized by the string tension $1/\alpha' = \tilde{R} / l_p^3$. In this case, the dimensionless quantity that we would like to hold fixed is the tension in units of the radius of the sphere, given by 
\be \label{tensu}
 { r^2 \over \alpha'} = \tilde R r^2 =  { R \over \mu r} r^2 =  
{\left( g_0^2 N  \right)^{1/4} \over \mu^{3 / 4}}
\ee
where we used (\ref{radius}).\footnote{Note that we may also write this as $\lambda^{1/4}$ where $\lambda = N^4/(\mu p^+)^3$ was the parameter (\ref{coupling}) controlling perturbation theory about the $X=0$ vacuum.} 
Since we are holding $\mu$ fixed, the NS5 limit corresponds to the 't Hooft limit of the matrix model. The strings of the little string theory are 
the usual 't Hooft strings. 
This makes this discussion be very similar to the discussion 
of Polchinski and Strassler \cite{jpms} for D3 branes. 

The limit that defines the M5 theory is the limit where, in 
addition, we take (\ref{tensu}) to infinity. 
This is a definition of the M5 brane theory which preserves 
16 of its supersymmetries in an explicit way. 

\vskip 1cm

\noindent
{\large \bf Acknowledgments}
\vskip .4cm
We would like to thank Keshav Dasgupta and Rob Myers for helpful discussions.
The work of M. M. Sh-J. and M.V.R is supported in part by NSF grant
PHY-9870115 and in part by funds from the Stanford Institute for
Theoretical Physics. The work of M.V.R. is also supported by NSERC
grant 22R81136 and by the Canada Research Chairs programme.
 The work of J. M. is supported by DOE grant DE-FG02-90ER40542.

\vskip 1cm

\appendix

\section{Protected states in the matrix theory spectrum}

In this Appendix we present a simple argument, without much group
theory, for why some states are protected. A full discussion can be
found in \cite{dsv2}.
The idea is to define an index, as in \cite{wittenindex}, to which
only the states in question contribute at weak coupling.
The index will be independent of the coupling so its value can
be computed at weak coupling.
Below we explain this in more detail.

It will be important for us to concentrate on the bosonic symmetry
generators that we can simultaneously diagonalize, $H$, $M^{12}$,
$M^{45} $, $M^{67}$, $M^{89}$. These are the Hamiltonian of the matrix
model, a $U(1)$ subgroup of the  $SU(2)$ symmetry rotating the first three
coordinates and a $U(1)^3$ subgroup of the $SO(6)$ rotating the other
six coordinates.  The supercharges transform as spinors  under
the rotation groups and they raise or lower the value of the
matrix model energy. It is convenient to focus on the energy raising
supercharge
$ Q^\dagger  =  Q^\dagger_{- +++}$,
where the subindices indicate the transformation
properties under  $M^{12},
M^{45},M^{67}, M^{89}$. This supercharge and its adjoint obey the
anticommutation relation
\be  \label{acmrel}
\{ Q, Q^\dagger \} = H - { 1 \over 3} M^{12} - { 1 \over 6 } (
M^{45}+ M^{67} +  M^{89})  \equiv \tilde H
\ee
where we have defined $\tilde H $. Note that $\tilde H$ has
a non-negative spectrum.  We should also note that $\tilde H$
commutes with $Q, Q^\dagger $. We could then consider the index
$Tr [ (-1)^F e^{ - \beta \tilde H} ]$ which receives
contributions only from  BPS states with $\tilde H =0$.
We find however that in perturbation theory there are many bosonic
and fermionic states with $\tilde H =0$. For this reason it is
convenient to introduce  another operator $J$ which commutes with
$Q, Q^\dagger $ (and therefore with $\tilde H$) and restrict the
index to subspaces with definite values of $J$.
It is convenient to pick  $J = H - { 1 \over 6 } M^{89}$.
Then one can check that 
all states with $\tilde H =0$ have $ J \geq 0$, and the only
perturbative states with $J=0$ are the ones created by
an oscillator mode in the $89$ plane with positive angular momentum
$M^{89}$.
In the notation of section 2 this is $A_8^\dagger  + i A_9^\dagger$.
Since all states created by this mode are bosonic we do not have any
cancellations in the index. So all BPS states with $\tilde H =0$ and
$J =0$ are exactly protected in the full theory.
Of course we should only consider gauge invariant states.
Once we take into account the $SO(6)$ symmetry we see that we recover
the result stated in (\ref{protected}) and used in section 2.

\section{Linear fluctuation spectrum of the fivebrane}

In this Appendix, we compute the linear fluctuation spectrum for the spherical
vacuum state of an M-theory fivebrane in the maximally supersymmetric plane wave
background of eleven-dimensional supergravity, given by
\bea\label{PPwave}
ds^2 &=& - 2 dx^+ dx^- + \sum_{A=1}^9 dx^A dx^A - \left(\sum_{i=1}^3 {\mu^2
\over 9} x^i x^i + \sum_{a=4}^9 {\mu^2 \over 36} x^a x^a\right) dx^+ dx^+\cr
F_{123+} &=& \mu
\eea

Ignoring for now the worldvolume two-form field and fermions, we find that the
light-cone gauge Hamiltonian for a fivebrane in this background is given by
\beas
{\cal H} &=& \int d^5 \sigma {1 \over 2 p^+} \{ P_A^2 +|g_{AB}| \} - {p^+ \over
2}
 g_{++} (X) - i_+ C^{(6)} \\
&=& {1 \over 2 p^+}  \left( P_A^2  + {1 \over 5!} \{ X^{A_1}, \dots, X^{A_p} \}
\{ X^{A_1}, \dots, X^{A_p} \} \right)\\
&& + {p^+ \over 2} \left( \left( {\mu \over 6} \right)^2 X^a X^a + \left( {\mu
\over 3} \right)^2 X^i X^i \right)   + {\mu \over 6!} \epsilon_{a_1 \cdots a_6}
X^{a_1} \{X^{a_2}, \dots, X^{a_6} \}
\eeas
where we define
\[
\{ A_1, \dots, A_5 \} = \epsilon^{\alpha_1, \dots, \alpha_5} \partial_{\alpha_1}
A_1 \cdots \partial_{\alpha_5} A_5
\]
and we have chosen worldvolume coordinates $\sigma_\alpha$ such that $d \sigma_1
\cdots d \sigma_5$ is the volume element.

To see that the spherical fivebrane represents a zero-energy solution, note that
setting $X^i=0$, the potential may be written as a perfect square,
\be
\label{square}
V_{X^a} = {1 \over 2 p^+} \left( {\mu p^+ \over 6} X^a + {1 \over 5!}
\epsilon^{a a_1 \cdots
a_5} \{ X^{a_1}, \dots, X^{a_5} \} \right)^2
\ee
It is convenient to define functions $x^a(\sigma)$ which map the worldvolume
into a target space unit sphere. Then $x^a x^a = 1$ and
\be\label{5sphere}
\{x^{a_1}, \dots, x^{a_5} \}= \epsilon^{a_1 \cdots a_6} x^{a_6}
\ee
From this relation, it is clear that
\[
X^a = r x^a
\]
gives a zero-energy solution if the sphere radius satisfies
\[
r^4 =  {\mu p^+ \over 6} \; .
\]

We would now like to expand the potential about this classical solution and
determine the quadratic fluctuation spectrum.

\subsection*{$X^a$ fluctuations}

Setting $X^a = r x^a + Y^a$, we find from (\ref{square}) that the quadratic
potential for the $X^a$ fluctuations is
\[
V_2^{X^a} = \left( {p^+ \over 2} \right) \left( {\mu \over 6} \right)^2 \left(
Y^a + {1 \over 24}
\epsilon^{a a_1 \cdots a_5} \{ x^{a_1}, \dots, x^{a_4}, Y^{a_5} \} \right)^2
\]
Normal modes will be solutions of the eigenvalue equation
\be\label{Ya}
{\cal L}_{ab} Y_b \equiv {1\over 24} \epsilon^{a a_1 \cdots a_5}
\{ x^{a_1}, \dots, x^{a_4}, Y^{a_5} \} = \lambda Y^a
\ee
with with masses given by
\[
M^2 = \left( {\mu \over 6}(1+\lambda) \right)^2 \; .
\]
Here, ${\cal L}_{ab}$ are generators of $SO(6)$, so the eigenvectors
will be (vector) spherical harmonics of $SO(6)$ given explicitly by
\beas
\ba{ll}
Y_l^a = S_{a a_1 \cdots a_l} x^{a_1} \cdots x^{a_l} &\qquad \qquad M
= {\mu \over 6} (l+1) \\
\tilde{Y}_l^a = x^a \tilde{S}_{a_1 \cdots a_{l-1}} x^{a_1} \cdots
x^{a_{l-1}} - {l \over 2l+2} \tilde{S}_{a a_1 \cdots a_{l-2}} x^{a_1}
\cdots x^{a_{l-2}} &\qquad \qquad M = {\mu \over 6} (l+3)\\
\hat{Y}_l^a = A^a_{a_1 \cdots a_l} x^{a_1} \cdots x^{a_l} &\qquad \qquad M = 0
\ea
\eeas
Here the tensors $S$ and $\tilde{S}$ are symmetric and traceless and $A$ is an
$SO(6)$ tensor with indices $a$ and $a_1$ antisymmetric. These correspond to
the three irreducible representations in the tensor product of the vector and
l-index symmetric traceless representations of $SO(6)$, with Dynkin labels
$(0,l+1,0)$, $(0,l-1,0)$ and $(1,l-1,1)$ in the order listed above.
The representations and energies of the modes $Y$ and $\tilde{Y}$
match exactly with the representations {\bf a} and {\bf f} of figure 1, with
$n=l+1$. The zero-modes $\hat{Y}$ are non-physical since they are
fluctuations in the gauge orbit directions under the gauge group of
volume-preserving diffeomorphisms.

\subsection*{$X^i$ fluctuations}

For the $X^i$ modes, we find that the quadratic action is given by
\beas
S_2^{X^i} = {p^+ \over 2} \int \left( {\mu \over 6} \right)^2 \left(4 X^i X^i + {1
\over 24} \{ x^{a_1}, \dots, x^{a_4}, X^i \}^2 \right)
= {p^+ \over 2} \left( {\mu \over 6} \right)^2 \int X^i (4+ {\cal L}_{ab}{\cal
L}^{ba}) X^i\ .
\eeas
Thus, again the eigenstates will be spherical harmonics on $S^5$,
given explicitly as symmetric traceless polynomials
\[
X^i_l = S^i_{a_1 \cdots a_l} x^{a_1} \cdots x^{a_l}
\]
with corresponding masses
\be
M^2=\left( {\mu \over 6} \right)^2
\left[l(l+4)+4\right]=\left( {\mu \over 6} (l+2) \right)^2\ .
\ee
Thus, we get a set of states which are vectors of $SO(3)$ and
l-index symmetric traceless tensors of $SO(6)$ with energies ${\mu
\over 6} (l+2)$. These match exactly with the representations {\bf
c} in figure 1.

\subsection*{Fermion fluctuations}

The quadratic potential for fermions may be determined just as for the case of
the supermembrane in section 2 of \cite{dsv1}. We start from the superspace M5-brane action in a form valid for coset spaces \cite{claus}, insert the component
field expressions for the superfields (known to all orders for coset spaces),
and choose the gauge $\Gamma^+ \theta = 0$. In this way, we find a
quadratic fermion potential given by
\[
V^\psi = -{i \over 8} \mu p^+ \Psi^T \gamma^{123} \Psi + {i \over 48} \Psi^T
\gamma^{ABCD} \{X^A, X^B, X^C, X^D, \Psi \}
\]
Expanding about the spherical fivebrane solution $X^a = r x^a$, $X^i=0$, we
obtain\footnote{The conventions used here for fermions are described
in Appendix A of \cite{dsv1}. In particular, $I,J$ and $\alpha,\beta$
are $SU(4)$ and $SU(2)$ indices respectively.}
\beas
V^\psi &=& - {i \over 8} \mu p^+ \Psi^T \gamma^{123} \Psi + {i \mu \over 48}
r^4 \Psi^T \gamma^{abcd} \{x^a, x^b, x^c, x^d, \Psi \}\\
&=& {\mu \over 4} \psi^{\dagger I \alpha} \psi_{I \alpha} - {\mu \over 12}
\psi^{\dagger I \alpha} g^{ab}_{\ I} {}^J {\cal L}_{ab} \psi_{J \alpha}\ ,
\eeas
The normal modes will be eigenstates of the equation
\[
g^{ab}_{\ I} {}^J {\cal L}_{ab} \psi_{J \alpha} = \lambda \psi_{I \alpha}
\]
with frequencies given by
\be
\label{fermmass}
\omega = {\mu \over 4} - {\mu \over 12} \lambda \; .
\ee
Again, these are given in terms of symmetric traceless
polynomials in the $x^a$,
\beas
\psi^l_{I} &=& (\theta_{I a_1 \cdots a_l} + g^{b a_1}_I {}^J \theta_{J b a_2
\cdots
a_l}) x^{a_1} \cdots x^{a_l} \qquad \qquad \lambda = -l\\
\tilde{\psi}^l_{I} &=& (l\theta_{I a_1 \cdots a_l} + (l+4) g^{a_1 b}_I {}^J
\theta_{J
b a_2 \cdots a_l}) x^{a_1} \cdots x^{a_l} \qquad \qquad \lambda = l+4\\
\eeas
where $\theta$ is totally symmetric and traceless in its $SO(6)$ indices and
we have suppressed the $SU(2)$ index.

The two eigenmodes correspond to the two irreducible representations of
$SO(6)$ obtained from the tensor product of the symmetric-traceless l-index
tensor with
a spinor. The modes $\psi^l$, with energy ${\mu \over 6}(l+{3 \over 2})$
correspond to the irreps with $SU(4)$ Dynkin labels $(1,l,0)$, and these match
precisely with the representations {\bf b} in figure 1 (where n=l+1). The
modes
$\tilde{\psi}^l$, with negative frequency $\omega = -{\mu \over 6}(l + {1
\over
2})$ correspond to the irrep with Dynkin label $(0,l-1,1)$. To compare with
the
matrix model spectrum, we should consider the positive-frequency complex
conjugate modes, which have energy ${\mu \over 6}(l + {1 \over 2})$ and lie in
the representations with Dynkin label $(1,l-1,0)$. These match exactly with
the representations {\bf e} in figure 1.

\subsection*{Two-form fluctuations}

To determine the two-form field fluctuations, we may begin directly with the
equation of motion for the two-form field $b$ in a general background, given as
equation (9) in \cite{pst}. Expanding to quadratic order, we have
\[
F_{qrs} g^{sp} \partial_p a = {1 \over 6 \sqrt{-g}} g_{qm} g_{rn}
\epsilon^{mnlrsp} \partial_l a F_{rsp}
\]
where
\[
F_{rsp} = 3 \partial_{[r} b_{sp]}
\]
Here $a$ is the auxiliary PST scalar and indices $p,q,r,\dots = 0,
\dots, 5$ are covariant worldvolume indices.
We may use the gauge symmetries to fix $a = \tau$ and $b_{0 \alpha} = 0$
($\alpha=1,\cdots, 5$). Then the equation of motion becomes
\[
\partial_0 b_{\alpha \beta} = {1 \over 2 g^{00} \sqrt{-g}} g_{\alpha \rho}
g_{\beta \sigma} \epsilon^{\rho \sigma \mu \nu \lambda} \partial_\mu b_{\nu
\lambda}
\]
To find the normal modes, we set
\[
b_{\alpha \beta} = e^{i \omega t} e_\alpha^a e_\beta^b B_{ab}(x)
\]
where $e_\alpha^a \equiv \partial_\alpha x^a$ and $B_{ab}$ may be chosen to
satisfy $x^a B_{ab} = 0$. Then the equation of motion gives the eigenvalue
equation
\[
{\mu \over 12} \epsilon^{\mu \nu \rho \sigma \tau} e_\mu^a e_\nu^b e_\rho^c
e_\sigma^d \partial_\tau B_{cd}  \equiv  {\mu\over 12} \epsilon^{abcdef} {\cal L}_{cd} B_{ef} = i \omega B_{ab}
\]
The normal modes may be determined by expanding $B_{ab}$ in terms of traceless
symmetric polynomials in $x^a$ and diagonalizing the resulting equation.

We define
\beas
B^l_{ab} &=& (B_{ab;a_1 \cdots a_l} - x^a x^c B_{cb;a_1 \cdots a_l} + x^b x^c
B_{ca;a_1 \cdots a_l})x^{a_1} \cdots x^{a_l}\\
\tilde{B}^l_{ab} &=& \epsilon_{abcdef} B_{cd ; e a_2 \cdots a_l} x^{a_2}
\cdots
x^{a_l}\\
\hat{B}^l_{ab} &=& (B_{a a_1;b a_2 \cdots a_n} - B_{b a_1 ; a a_2 \cdots a_n}
+ x^a x^c B_{b a_1;c a_2 \cdots a_n} - x^b x^c B_{a a_1 ;c a_2 \cdots a_n})
x^{a_1} \cdots x^{a_n}\\
\eeas
where $B_{ab;a_1 \cdots a_l}$ is a traceless tensor antisymmetric in $a,b$ and
symmetric in $a_1, \dots ,a_l$. Then the eigenmodes are given by
\beas
\ba{ll}
B^{l \pm}_{ab} = \mp {in \over 2} B^l_{ab} + (n+2) \tilde{B}^l_{ab} \pm i
\hat{B}^l_{ab} &\qquad \qquad \omega = \pm {\mu \over 6} (l+2) \\
B^{l 0}_{ab} = B^l_{ab} + \hat{B}^l_{ab} &\qquad \qquad \omega = 0
\ea
\eeas
These normal modes correspond to the three irreducible representations in the
traceless $SO(6)$ tensor product of the l-index symmetric tensor with the 2-
index antisymmetric tensor. The modes $B^{l \pm}$, with energy ${\mu \over
6}(l+2)$ lie in the $SO(6)$ representations with Dynkin labels $(2,l-1,0)$ and
$(0,l-1,2)$.
They comprise the positive and negative frequency parts of a set
of modes matching exactly with the matrix theory modes {\bf d} in figure 1.
The zero-modes correspond to the representation with Dynkin label $(1,l,1)$,
but these are non-physical since they correspond to gauge variations with
gauge transformation parameter
\[
\Lambda_\alpha = e_\alpha^a B_{ab; a_1 \cdots a_n} x^b x^{a_1} \cdots x^{a_n}
\; .
\]

\section{Decoupled 2+1 dimensional field theory from the matrix model}

In this Appendix, we determine explicitly the form of the
commutative 2+1 dimensional field theory arising from the matrix
model action expanded about the $k$-membrane vacuum in the limit
where $N \to \infty$ with $(R / \mu)^3/N$ fixed.

We begin by rescaling things so that everything is dimensionless
and the quadratic action is independent of the parameters. Then in
the limit, we get a continuum theory on a sphere (which we
initially take to be a unit sphere) via the following
replacements: \beas
{\rm Tr} & \to & \int d \Omega\ {\rm tr}\\
M &\to & M(\theta, \phi)\\
{}[ J^i, M ] & \to & L_i M \qquad \qquad L_i \equiv -i
\epsilon^{ijk} x^j
\partial_k\\
{}[ A,B ] & \to & {1 \over \sqrt{N}}[A,B]\ ,
\eeas
where the objects on the right hand side are 
$k\times k$ matrices.
Here and below,
$x^i$ are the embedding coordinates of the unit sphere, which
satisfy $x^2 = 1$. With these replacements, the action becomes
\beas
S &=& \int dt d \Omega\ {\rm tr} \left( {1 \over 2} \dot{X}^a
\dot{X}^a + {1 \over 2} \dot{Y}^i \dot{Y}^i + {i \over 2}
\Psi^\dagger \dot{\Psi}
\right.\\
&& -{1 \over 2} {1 \over 36} X^a X^a + {1 \over 2} {1 \over 9}
({\cal L}_i X^a)^2 - {1 \over 2} {1 \over 9} ({\cal Y}_i)^2 - {i
\over 8}
\Psi^\dagger \gamma^{123} \Psi\\
&& \left. {1 \over 6} \Psi^\dagger \gamma^i {\cal L}_i \Psi + {g
\over 6} \Psi^\dagger \gamma^a [X^a, \Psi] + {1 \over 36} g^2
[X^a,X^b]^2 \right)
\eeas
Here, we have defined
\beas
g &=& ({R \over \mu})^{3 \over 2}{3 \over \sqrt{N}}\\
{\cal L}_i A &=& L_i A  + g [Y^i, A]\\
{\cal Y}_i &=& Y_i + {i \over 2} \epsilon^{ijk} (L_j Y_k - L_k Y_j
+ g [Y_j, Y_k])
\eeas
The gauge symmetry transforms the fields as
\beas
\delta Y_i &=& L_i \lambda + g [Y^i, \lambda]\\
\delta X_a &=& g [X_a, \lambda]\\
\delta \Psi &=& g [\Psi, \lambda]
\eeas
and ${\cal L}$ and ${\cal
Y}$ are defined to be covariant. The supersymmetry transformation
rules are
\beas
\delta X_a &=& -i \epsilon \gamma^a \Psi \\
\delta Y^i &=& -i \epsilon \gamma^i \Psi\\
\delta \Psi &=& \dot{X}^a \gamma^a \epsilon + \dot{Y}^i \gamma^i
\epsilon - {1 \over 6} X^a \gamma^a \gamma^{123} \epsilon\\
&& + {i \over 6} g [X^a, X^b] \gamma^{ab} \epsilon + {i \over 3}
{\cal L}_i X^a \gamma^{ia} \epsilon + {1 \over 6} \epsilon^{ijk}
{\cal Y}^i \gamma^{jk} \epsilon \eeas The action may then be
rewritten by splitting
\[
Y^i = x^i \phi + \epsilon^{ijk} x^j A_k
\]
It may be checked that
\[
A_i = \epsilon^{ijk} Y_j x^k
\]
transforms like a conventional gauge field while
\[
\phi = x^i Y^i
\]
transforms like an adjoint scalar. Note that though $A$ has three
components in this notation, these always point tangent to the
sphere, so we could rewrite $A$ in terms of a two-component vector
field with a worldvolume index. Rescaling the bosonic fields by
$\sqrt{3}$, the coupling $g$ by $1/ \sqrt{3}$ and the time $t$ by
3 (for convenience), and reintroducing $A_0$ the resulting action
becomes:
\beas
S &=& \int dt d \Omega\ {\rm tr} \left( -{1 \over 4} F^{\mu
\nu} F_{\mu \nu} - {1 \over 2} (D_\mu X^a)^2 - {1 \over 2} (D_\mu
\phi)^2 +{i \over 2} \Psi^\dagger D_0 \Psi - {i \over 2}
\epsilon^{ijk}
\Psi^\dagger \gamma^i x^j D_k \Psi \right.\\
&& + {g \over 2} \Psi^\dagger \gamma^i x^i [\phi, \Psi] + {g \over
2} \Psi^\dagger \gamma^a [ X^a, \Psi]
 + {1 \over 4} g^2 [X_a, X_b]^2 + {1 \over 2} g^2 [\phi, X^a]^2 \\
&& \left. -{1 \over 8} X^a X^a - {1 \over 2} \phi^2 - {3i \over 8}
\Psi^\dagger \gamma^{123} \Psi + {1 \over 2} \phi \epsilon^{ijk}
x^i F_{jk} \right)
\eeas
Here, the first and second lines are just
the usual 2+1 dimensional SYM theory. The first line contains the
standard kinetic terms for the fields, the second term contains
the usual D2-brane interactions, and the final line contains
masses for the scalars and fermions and an extra $\phi F$
interaction. The supersymmetry transformation rules may be
obtained from the ones above by substituting for $Y$. Rescaling this action to make the worldvolume a sphere of radius $1/\mu$ gives the desired D2-brane theory (\ref{dtwotheory}) above.

\section{ Relationship to the Polchinski-Strassler discussion of
$ { \cal N} = 1^* $ theories. }

The picture we have presented for the matrix model
was inspired by a similar discussion for D3 branes in the context
of ${ \cal N} =1^*$ theories in \cite{jpms} .
Namely, \cite{jpms}
considered  ${ \cal N} =4$ Yang-Mills in four dimensions
and added a quadratic superpotential that gave a mass to the three
chiral multiplets. For weak gauge coupling the vacua are labelled
by $SU(2)$ representations. These can be interpreted as D5 branes
wrapping an $S^2 \times R^4$.
 At strong coupling the vacuum with $\phi=0$ (the
trivial $SU(2)$ representation) should be thought of in terms
of an expanded spherical NS5-brane. This vacuum is related by
S-duality to the vacuum with a single D5 brane.

Here we just point out that this ${ \cal N} =1^*$ arises as the
DLCQ theory of M-theory on $T^3$ with a plane wave in the 8 non-compact
directions.
More explicitly, we can consider the following background in
8 noncompact dimensions
\be \label{psba}
 \ba{r l} 
ds^2 =& - 2 dx^+ dx^- - | \partial W|^2 (dx^+)^2 + dz_i d \bar z_i
 \\
F_4  = & dx^+ \partial_i \partial_ j W  \eta^{i \bar i} \epsilon_{
\bar i \bar l \bar m} d{\bar z}^{\bar l} d{\bar z}^{\bar m} dz^j   + c.c.
\ea
\ee
Where $W$ is an arbitrary holomorphic function of the three
complex coordinates $z_i$, $i =1,2,3$.
This background is related by T and U dualities to the backgrounds
considered in \cite{jmlm}.
The DLCQ version of this background, with $x^-$ compactified,
is expected to be described by ${ \cal N} =4$ Yang Mills theory
with a superpotential given by $W$ and compactified on $T^3$.\footnote{
For a non-renormalizable superpotential this  DLCQ
description is not well defined. }
For a quadratic superpotential we obtain the ${ \cal N} =1^*$ of
\cite{jpms}.
By performing U-dualities on the background described above 
(\ref{psba}) one
can obtain backgrounds such that when we put D3 branes on them
we get an arbitrary superpotential $W$ on the D3  worldvolume theory.
These backgrounds were also studied in \cite{{jpmg},{mg}}.

In the context of the DLCQ description of M-theory on $T^3$, the
fact that the transverse M5 is related to an NS5-brane was
used in \cite{grt} to give some insight on the problem of the
transverse M5 branes.

\end{document}